\documentclass[amsmath,amssymb]{revtex4}
\newtheorem{theorem}{Theorem}
\newtheorem{corollary}{Corollary}

\usepackage{rotating}

\begin{document}
\title{Solvable Systems of Linear Differential Equations}

\author{Katherine M. Robertson}%
\author{Nasser Saad}
\email{kmrobertson@upei.ca}
\email{nsaad@upei.ca}
\affiliation{Department of Mathematics and Statistics, University of Prince Edward Island 
Charlottetown, Prince Edward Island C1A 4P3, Canada}%
%
\def\dbox#1{\hbox{\vrule  
                        \vbox{\hrule \vskip #1
                             \hbox{\hskip #1
                                 \vbox{\hsize=#1}%
                              \hskip #1}%
                         \vskip #1 \hrule}%
                      \vrule}}
\def\qed{\hfill \dbox{0.05true in}}  
\def\square{\dbox{0.02true in}} 
\begin{abstract}
\noindent{\bf Abstract:}~~The asymptotic iteration method (AIM) is an iterative technique used to find exact and approximate solutions to second-order linear differential equations.  In this work, we employed AIM to solve systems of two first-order linear differential equations.  The termination criteria of AIM will be re-examined and the whole theory is re-worked in order to fit this new application. As a result of our investigation, an interesting connection between the solution of  linear systems and the solution of Riccati equations is established. Further, new classes of exactly solvable systems of linear differential equations with variable coefficients are obtained. The method discussed allow to construct many solvable classes through a simple procedure.
\end{abstract}
\vskip0.1 true in
\maketitle
   \vskip0.1true in                        
\noindent{\bf keywords:} {~Asymptotic iteration method, Linear systems, Coupled differential equations, Decoupled differential equations, Riccati equation, Confluent Hypergeometric function, Gaussian Hypergeometric function, Hermite polynomials.
}


\section{Introduction} 
\noindent There are a limited number of exactly solvable systems of differential equations that are known in the literature \cite{polyanin}-\cite{kamke}.  For example, the system of two constant-coefficient differential equations:
\begin{align}
\phi_{1}'=& a\phi_{1} + b\phi_{2} \notag \\
\phi_{2}'=& c\phi_{1} + d\phi_{2} \label{1}
\end{align}
has a general solution depends on the roots of the characteristic equation \cite{polyanin}-\cite{kamke}:
\begin{equation} \label{2}
\lambda^{2} - (a+d)\lambda + (ad-bc) = 0. 
\end{equation}
The purpose of the present work is two-fold: (1) to introduce a new technique based on the asymptotic iteration method \cite{hakan1}-\cite{hakan4} to analyze the exact solutions of linear systems
\begin{align}
\phi_{1}'=& \lambda_0(x)\phi_{1} + s_0(x)\phi_{2}, \notag \\
\phi_{2}'=& \omega_0(x)\phi_{1} + \rho_0(x)\phi_{2}, \label{3} 
\end{align}
where $\lambda_0(x)$, $s_0(x)$, $\omega_0(x)$ and $\rho_0(x)$ are continuous and differentiable functions, and (2) to add new classes of exactly solvable systems to the known ones \cite{polyanin}-\cite{kamke}. 
\vskip0.1true in 
The asymptotic iteration method (AIM) is an iterative technique introduced  originally \cite{hakan1} to find the exact and approximate solutions of second-order linear homogeneous differential equations \cite{hakan3}.  Since then, the method has enjoyed a number of interesting applications, particularly, in relativistic and non-relativistic quantum mechanics \cite{hakan1}-\cite{ciftci} . In their study of Dirac equation, Hakan et al \cite{hakan2} modified AIM to study Dirac equation by analyzing a system of linear differential equations, however, their work was limited to this particular application. In the present work, we extend AIM to deal with different systems of first-order linear differential equations in general, and to give a detail structure useful in generating exactly solvable systems.  
\vskip0.1true in
The paper is organized as follows: in next section, we re-examine AIM to deal with a system of linear differential equations.  In section III, we used AIM to provide a simple recipe to find the exact solution of a system of two constant-coefficient differential equations (\ref{1}). In section IV, we give a necessary condition for polynomial solutions of linear systems of differential equations (\ref{3}). In section V, we establish an interesting connection between the exact solution of linear system (\ref{3}) and the exact solution of a Riccati equation; namely, for every exactly solvable Riccati equation, there exist at least one system of linear differential equations  that is exactly solvable. In section VI, some applications and tables of results are reported. The conclusion is given in section VII. 

\section{Iterative solution of Linear systems} 
\noindent We consider a system given by (\ref{3}), direct differentiation with respect to $x$ yields
\begin{align}
\phi_{1}'' =& \lambda_{1}\phi_{1} + s_{1}\phi_{2}  \notag\\
\phi_{2}'' =& \omega_{1}\phi_{1}+\rho_{1}\phi_{2} \label{4}
\end{align}
where
\begin{align}
\lambda_1 =&~\lambda_{0}' +\lambda_{0}^{2} +s_{0}\omega_{0}\notag\\
s_1=&~s_{0}'+\lambda_{0}s_{0}+s_{0}\rho_{0} \notag\\
\omega_1=&~\omega_{0}' +\omega_{0}\lambda_{0}+\rho_{0}\omega_{0}\notag\\
\rho_1=&~\rho_{0}'+\omega_{0}s_{0}+\rho_{0}^{2}\label{5}
\end{align}
In general, the $n^{th}$ and $(n+1)^{th}$ derivative of (\ref{4}) are:
\begin{align} 
\phi_{1}^{(n+1)} = \lambda_{n}\phi_{1} + s_{n}\phi_{2}\notag \\ 
\phi_{2}^{(n+1)} = \omega_{n}\phi_{1} + \rho_{n}\phi_{2} \label{6}
\end{align}
and
\begin{align} 
\phi_{1}^{(n+2)} = \lambda_{n+1}\phi_{1} + s_{n+1}\phi_{2} \notag \\ 
\phi_{2}^{(n+2)} = \omega_{n+1}\phi_{1} + \rho_{n+1}\phi_{2} \label{7}
\end{align}
respectively, where $\lambda_{n+1}$, $s_{n+1}$, $\omega_{n+1}$ and $\rho_{n+1}$ are computed recursively  by:
\begin{align}  
\lambda_{n+1} &= \lambda_{n}' + \lambda_{n}\lambda_{0} + s_{n}\omega_{0}, \notag \\ 
s_{n+1} &= s_{n}' + \lambda_{n}s_{0} + s_{n}\rho_{0}, \notag \\ 
\omega_{n+1} &= \omega_{n}' + \omega_{n}\lambda_{0} + \rho_{n}\omega_{0},  \notag \\ 
\rho_{n+1} &= \rho_{n}' + \omega_{n}s_{0} + \rho_{n}\rho_{0}. \label{8} 
\end{align}
The ratio $\phi_{1}^{(n+2)}/\phi_{1}^{(n+1)}$ then reads 
\begin{equation} \label{9}
\frac{d}{dx}\ln(\phi_{1}^{(n+1)})=\frac{\lambda_{n+1}(\phi_{1}+\frac{s_{n+1}}{\lambda_{n+1}}\phi_{2})}{\lambda_{n}(\phi_{1}+\frac{s_{n}}{\lambda_{n}}\phi_{2})}.
\end{equation}
If, for sufficiently large $n>0$, we have
\begin{equation} \label{10}
\frac{s_{n}}{\lambda_{n}} = \frac{s_{n+1}}{\lambda_{n+1}} \equiv \alpha
\end{equation}
then (\ref{9}) reads
\begin{equation} \label{11}
\frac{d}{dx}\ln(\phi_{1}^{(n+1)})=\frac{\lambda_{n+1}}{\lambda_{n}}
\end{equation}
and consequently, we have
\begin{equation} \label{12}
\phi_{1}^{(n+1)} = C_{1}\exp\left( \int^{x} \frac{\lambda_{n+1}}{\lambda_{n}} dt \right) = C_{1}\lambda_{n}\exp\left(\int^{x}\left(\alpha\omega_{0} + \lambda_{0}\right) dt\right)
\end{equation}
where we used (\ref{8}) and (\ref{10}). Further, using (\ref{6}), we have 
\begin{equation} \label{13}
\phi_{1} + \alpha \phi_{2} = C_{1}\exp\left( \int^{x} \left( \alpha\omega_{0} +\lambda_{0} \right) dt \right)
\end{equation}
which can be solve for $\phi_1$ to yields, using (\ref{3}), the first-order differential equation of $\phi_2$ 
\begin{equation} \label{14}
\phi_{2}' + \left(\omega_{0}\alpha - \rho_{0}\right)\phi_{2} = \omega_{0}C_{1}\exp\left(\int\left(\omega_{0}\alpha+\lambda_{0}\right)dx\right).
\end{equation}
The exact solution of this first-order differential equation is easily obtained; namely
\begin{align} 
\phi_{2} =& \exp\left( \int^{x} \left( \rho_{0} - \omega_{0}\alpha \right) dt \right)\left[ C_{2} + C_{1} \int^{x} \left( \omega_{0} \exp\left( \int^{t} \left( \lambda_{0} - \rho_{0} + 2\omega_{0}\alpha \right) d\tau \right) \right) dt \right] \label{15}
\end{align}
where $C_1$ and $C_2$ are the constant of integration. Consequently, we have
\begin{equation}\label{16}
\phi_1= C_{1}\exp\left( \int^{x} \left( \alpha\omega_{0} +\lambda_{0} \right) dt \right)- \alpha\phi_2.
\end{equation}
Equivalently, if we consider instead the ratio $\phi_{2}^{(n+2)}/\phi_{2}^{(n+1)}$ and for sufficiently large $n>0$, we have
\begin{equation} \label{17}
\frac{\omega_{n+1}}{\rho_{n+1}} = \frac{\omega_{n}}{\rho_{n}} \equiv \beta 
\end{equation}
that
\begin{align}
\phi_{1} =& \exp\left(\int\left(\lambda_{0}-s_{0}\beta\right)dx\right)\left[C_2^\prime+C_1^\prime\int\left(s_{0}\exp\left(\int\left(\rho_{0}-\lambda_{0}+2s_{0}\beta\right)dx\right)\right)dx\right] \label{18}
\end{align}
and
\begin{equation} \label{19}
\phi_{2} = -\beta\phi_{1} + C_1^\prime\exp\left(\int\left(\beta s_{0}+\rho_{0}\right)dx\right) 
\end{equation}
where again $C_1^\prime$ and $C_2^\prime$ are constant.
The above discussion establish the following theorem:
\vskip 0.1 true in
\begin{theorem} \label{th1} The general solution of the system:
\begin{align} 
\phi_{1}' =& \lambda_{0}(x)\phi_{1} + s_{0}(x)\phi_{2} \notag \\ 
\phi_{2}' =& \omega_{0}(x)\phi_{1} + \rho_{0}(x)\phi_{2} \label{20}
\end{align}
is given by:
\begin{align} 
\phi_{2}(x) =& \exp\left( \int^{x} \left( \rho_{0}(t) - \alpha \omega_{0}(t)\right) dt \right) \left[ C_{2} + C_{1} \int^{x} \omega_{0}(t) \exp\left( \int^{\tau} \left( \lambda_{0}({\tau}) - \rho_{0}({\tau}) + 2\omega_{0}({\tau})\alpha \right) d\tau \right) dt \right] \notag \\
\phi_{1}(x) =& C_{1}\exp \left(\int^{x}\left(\alpha\omega_{0}(t) + \lambda_{0}(t) \right) dt \right) - \alpha \phi_{2}(x) \label{21}
\end{align}
if for sufficiently large $n\geq 0$
\begin{equation} \label{22}
\alpha \equiv \frac{s_{n}}{\lambda_{n}} = \frac{s_{n+1}}{\lambda_{n+1}}
\end{equation}
Equivalently, if for some $n\geq 0$, 
\begin{equation} \label{23}
\beta \equiv \frac{\omega_{n}}{\rho_{n}} = \frac{\omega_{n+1}}{\rho_{n+1}}
\end{equation}
then the general solution is given by:
\begin{align}
\phi_{1}(x) =& \exp\left(\int^x\left(\lambda_{0}(t)-\beta s_{0}(t)\right)dt\right) \left[C_{2}^\prime+C_{1}^\prime\int^x\left(s_{0}(t)\exp\left(\int^t\left(\rho_{0}(\tau)-\lambda_{0}(\tau)+2s_{0}(\tau)\beta\right)d\tau\right)\right)dt\right], \notag \\
\phi_{2}(x) =& -\beta\phi_{1}(x) + C_{1}^\prime\exp\left(\int^x\left(\beta s_{0}(t)+\rho_{0}(t)\right)dt\right) \label{24}
\end{align}
where
\begin{align}  
\lambda_{n+1} &= \lambda_{n}' + \lambda_{n}\lambda_{0} + s_{n}\omega_{0}, \notag \\ 
s_{n+1} &= s_{n}' + \lambda_{n}s_{0} + s_{n}\rho_{0}, \label{25} \\
\omega_{n+1} &= \omega_{n}' + \omega_{n}\lambda_{0} + \rho_{n}\omega_{0},  \notag \\ 
\rho_{n+1} &= \rho_{n}' + \omega_{n}s_{0} + \rho_{n}\rho_{0}. \notag 
\end{align}\qed
\end{theorem}
\vskip 0.1 true in
\noindent Note that (\ref{22}) and (\ref{23}) can be written in more convenient way as
\begin{equation}\label{26}
\frac{s_{n+1}}{\lambda_{n+1}} = \frac{s_{n}}{\lambda_{n}} \Leftrightarrow \delta_{n+1} = \lambda_{n+1}s_{n}-\lambda_{n}s_{n+1}=0, 
\end{equation}
and
\begin{equation}\label{27}
\frac{\omega_{n+1}}{\rho_{n+1}} = \frac{\omega_{n}}{\rho_{n}} \Leftrightarrow \Delta_{n+1} = \omega_{n+1}\rho_{n}-\omega_{n}\rho_{n+1}=0.  
\end{equation}

\noindent{\bf Example 1:} Consider the following linear system of differential equations
\begin{align}
\phi_{1}' =& \frac{a}{x}\phi_{1} + \frac{b}{x}\phi_{2} \notag \\
\phi_{2}' =& \frac{c}{x}\phi_{2} + \frac{d}{x}\phi_{2} \label{28}
\end{align}
where $a$, $b$, $c$, and $d$ are arbitrary constants with $ad-cb\neq 0$. It is straightforward, using (\ref{26}), to show that
\begin{equation}\label{29}
 \delta_n=-{b\over x^{2n+1}}\prod_{m=0}^{n-1}(m^2-m(a+d)+ad-bc), \quad n=1,2,\dots
\end{equation}
Clearly, $\delta_n=0$, if $(n-1)^2-(n-1)(a+d)+ad-bc=0$ or $a={bc-(n-1)^2+(n-1)d\over d-n+1}$. Furthermore, we have, using (\ref{22}),  $\alpha={d-(n-1)\over c}$ and consequently the system, for $n=1,2,\dots$
\begin{align}
\phi_{1}' =& \left[{bc-(n-1)^2+(n-1)d\over (d-n+1)x}\right]\phi_{1} + \frac{b}{x}\phi_{2} \notag \\
\phi_{2}' =& \frac{c}{x}\phi_{2} + \frac{d}{x}\phi_{2} \label{30}
\end{align}
has the general solution
\begin{align}
\phi_{1} =& C_1\left[{bc\over bc+(1+d-n)^2}\right]x^{d+{bc\over 1+d-n}}+C_2\left[ {n-d-1\over c}\right]x^{n-1}\notag \\
\phi_{2} =& C_1\left[{c(1+d-n)\over bc+(1+d-n)^2}\right]x^{d+{bc\over 1+d-n}}+ C_2x^{n-1}\label{31}
\end{align}
for $n=1,2,\dots$. Similar cases are reported in Table 1. Further examples of exactly solvable systems are reported in Table II.

\begin{table}
\begin{center} 
\caption{Exactly solvable systems $\phi_{1}'= \lambda_0(x)\phi_{1} + s_0(x)\phi_{2},$ and $\phi_{2}'= \omega_0(x)\phi_{1} + \rho_0(x)\phi_{2}$, for $\lambda_0(x),s_0(x),\omega_0(x)$ and $\rho_0(x)$ arbitrary functions. Here, $a,b,c$, and $d$ are constants and $n=1,2,\dots$.} 
\begin{tabular}{|l|l|l|l|l|} 
\hline\hline 
 $\lambda_0(x)$ & $s_0(x)$ & $\omega_0(x)$ & $\rho_0(x)$ & Solution \\ [0.5ex]
\hline 
${bc-(n-1)^2+(n-1)d\over (d-n+1)x}$ & ${b\over x}$  & ${c\over x}$ & ${d\over x}$  &  $\phi_{1} = C_1[{bc\over bc+(1+d-n)^2}]x^{d+{bc\over 1+d-n}}+C_2[ {n-d-1\over c}]x^{n-1}$\\
~&~&~&~&$\phi_{2} = C_1[{c(1+d-n)\over bc+(1+d-n)^2}]x^{d+{bc\over 1+d-n}}+ C_2x^{n-1}$\\[1ex]
\hline 
${a\over x}$ & ${ad-(a+d)(n-1)+(n-1)^2\over cx}$  & ${c\over x}$ & ${d\over x}$ & $\phi_1=C_1[{1+a-n\over 2+a+d-2n}]x^{1+a+d-n}+C_2[{n-d-1\over c}]x^{n-1}$\\
~&~&~&~&$\phi_{2} = C_1[{c\over 2+a+d-2n}]x^{1+d+a-n}+ C_2x^{n-1}$\\[1ex]
\hline
${a\over x}$ & ${b\over x}$   & ${(d-n+1)(a-n+1)\over bx}$ & ${d\over x}$ & $\phi_1=C_1[{1+a-n\over 2+a+d-2n}]x^{1+a+d-n}+C_2[{b\over n-a-1}]x^{n-1}$\\
~&~&~&~&$\phi_{2} = C_1[{(1+a-n)(1+d-n)\over b(2+a+d-2n)}]x^{1+a+d-n}+ C_2x^{n-1}$\\[1ex]
\hline
${a\over x}$& ${b\over x}$   & ${c\over x}$ & ${bc+(n-1)a-(n-1)^2\over (a-n+1)x}$ & $\phi_1=C_1[{(1+a-n)^2\over a^2+bc-2(n-1)a+(n-1)^2}]x^{a+{bc\over 1+a-n}}+C_2[{b\over n-a-1}]x^{n-1}$\\
~&~&~&~&$\phi_{2} = C_1[{c(1+a-n)\over a^2+bc-2(n-1)a+(n-1)^2}]x^{a+{bc\over 1+a-n}}+C_2x^{n-1}$\\[1ex]
\hline 
\end{tabular}
\end{center}
\end{table}

\begin{table}
\begin{center} 
\caption{Exactly solvable systems (\ref{1}) for different $\lambda_0(x),s_0(x),\omega_0(x)$ and $\rho_0(x)$ using Theorem 1 where $a,b,c,d$, and $a_i,i=1,\dots,4$ are arbitrary real numbers. Here $n=1,2,\dots$ and  $E_a[z]$ is the exponential integral function $E_a[z]=\int_1^\infty {t^{-a}e^{-zt}}dt$.} 
\begin{tabular}{|l|l|l|l|l|} 
\hline\hline 
 $\lambda_0(x)$ & $s_0(x)$ & $\omega_0(x)$ & $\rho_0(x)$ & Solution \\ [0.5ex]
\hline 
$a+{cd+n-1\over x}$ & ${c\over x}$  & ${d\over x}$ & ${n\over x}$  &  $\phi_{1} = C_1[{x^{cd+n}e^{ax}\over cd+ax}-{cd x^{n-1}}\int{x^{cd}e^{ax}\over (cd+ax)^2}dx]-C_2 cx^{n-1}$,\\
~&~&~&~&$\phi_{2} =(cd+ax)x^{n-1}(C_1 d \int{{x^{cd}e^{ax}\over (cd+ax)^2}dx}+ C_2).$\\[1ex]
\hline 
$a+{b\over x}$ & ${b-n+1\over d~x}$  & ${d\over x}$ & ${n\over x}$ & $
\phi_1=C_1({x^{b+1}e^{ax}\over 1+b-n+ax}+(1+b-n)x^{n-1}\int{x^{b+1-n}e^{ax}\over (1+b-n+ax)^2}dx)+C_2{(1+b-n)x^{n-1}\over d}$,\\
~&~&~&~&$\phi_{2} = -x^{n-1}(1+b-n+ax)(C_1 d \int{x^{1+b-n}e^{ax}\over (1+b-n+ax)^2}dx+C_2).$\\[1ex]
\hline
$a+{b\over x}$ & ${c\over x}$   & ${(b-n+1)\over c~x}$ & ${n\over x}$ & $\phi_1=C_1[
{e^{ax}x^{b+1}\over (1+b-n+ax)}+ (1+b-n)x^{n-1}\int{x^{1+b-n}e^{ax}\over (1+b-n+ax)^2}dx)+cC_2 x^{n-1}$,\\
~&~&~&~&$\phi_{2} = -x^{n-1}(1+b-n+ax)[{C_1(1+b-n)\over c}\int{x^{1+b-n}e^{ax}\over (1+b-n+ax)^2}dx+C_2].$\\[1ex]
\hline
${a_2a_3\over a_4}+{{ca_3\over a_4}+n-1\over x}$& $a_2+{c\over x}$   & $a_3+{a_3\over a_4 x}$ & $a_4+{n\over x}$ & $\phi_1=C_1x^{{a_3c\over a_4}+n}[e^{{a_2a_3x\over a_4}+a_4x}+E_{-{a_3c\over a_4}}[-({a_2a_3\over a_4}+a_4)x]+a_4 xE_{-{a_4+a_3c\over a_4}}[-({a_2a_3\over a_4}+a_4)x]$\\
~&~&~&~&$\quad-{a_4\over a_3}C_2x^{n-1},$\\
~&~&~&~&$\phi_2=-{a_3C_1\over a_4}x^{{a_3c\over a_4}+n}(E_{-{a_3c\over a_4}}[-({a_2a_3\over a_4}+a_4)x]+a_4xE_{-{a_4+a_3c\over a_4}}[-({a_2a_3\over a_4}+a_4)x])+C_2x^{n-1}.$\\
\hline 
\end{tabular}
\end{center}
\end{table}

\section{solution of linear constant-coefficient systems}
\begin{theorem} \label{Th2}
The constant-coefficient first-order linear system
\begin{align}
\phi_{1}' =&  \lambda_{0}\phi_{1} + s_{0}\phi_{2} \notag \\
\phi_{2}' =&  \omega_{0}\phi_{1} + \rho_{0}\phi_{2} \label{32}
\end{align}
where $\lambda_0,s_0,\omega_0$ and $\rho_0$ are constants, has the general solution
\begin{align}
\phi_{1} =& C_{1}\left[1-\frac{\omega_{0}\alpha}{\lambda_{0}-\rho_{0}+2\omega_{0}\alpha}\right]e^{(\lambda_{0} + \alpha\omega_{0})x} - \alpha C_{2}e^{(\rho_{0}-\omega_{0}\alpha)x} \notag \\
\phi_{2} =& C_{1}\frac{\omega_{0}}{\lambda_{0}-\rho_{0}+2\omega_{0}\alpha}e^{(\lambda_{0}+\omega_{0}\alpha)x}+C_{2}e^{(\rho_{0}-\omega_{0}\alpha)x}   \label{33}
\end{align}
where $\alpha$ is given by:
\begin{equation} \label{34}
\alpha = \frac{(\rho_0-\lambda_{0})+ \sqrt{(\lambda_{0}-\rho_{0})^{2}+4\omega_{0}s_{0}}}{2\omega_{0}}\quad\quad\mbox{or}
\quad \alpha = \frac{(\rho_0-\lambda_{0}) - \sqrt{(\lambda_{0}-\rho_{0})^{2}+4\omega_{0}s_{0}}}{2\omega_{0}}.
\end{equation}
\end{theorem}
\vskip0.1true in
\noindent{\bf Proof:} From the AIM sequence (\ref{25}), we have for the constant-coefficient linear system that   
\begin{align}  \label{35}
\lambda_{n+1} &=\lambda_{n}\lambda_{0} + s_{n}\omega_{0}, \notag \\ 
s_{n+1} &=\lambda_{n}s_{0} + s_{n}\rho_{0},\\
\omega_{n+1} &=\omega_{n}\lambda_{0} + \rho_{n}\omega_{0},  \notag \\ 
\rho_{n+1} &= \omega_{n}s_{0} + \rho_{n}\rho_{0}. \notag 
\end{align}
Consequently, the ratio (\ref{22}) now reads
\begin{equation}
\alpha={s_0+\alpha\rho_0\over \lambda_0+\alpha\omega_0}
\end{equation}
which yield the quadratic equation
\begin{equation*}
\omega_0 \alpha^2 +(\lambda_0-\rho_0)\alpha-s_0=0
\end{equation*}
with solutions given by
$\alpha = \frac{(\rho_0-\lambda_{0})\pm \sqrt{(\lambda_{0}-\rho_{0})^{2}+4\omega_{0}s_{0}}}{2\omega_{0}}.
$~The exact solutions (\ref{33}), then, follows immediately from (\ref{21}).\qed  

\vskip0.1true in
\noindent\textbf{Example 2:}
Consider the following elementary system:
\begin{align}\label{37}
\phi_{1}' =& \phi_{1} + 2\phi_{2} \notag \\
\phi_{2}' =& 3\phi_{1} + 2\phi_{2}
\end{align}
we have, using (\ref{34}), that $\alpha = 1,-{2\over 3}$. Thus, for $\alpha=1$, the exact solution of the given system is given by
\begin{align}\label{38}
\phi_{1} =& \frac{2}{5}C_1e^{4x}-C_{2}e^{-x}  \notag \\
\phi_{2} =&\frac{3}{5}C_1e^{4x} + C_{2}e^{-x}
\end{align}
while for $\alpha=-{2\over 3}$, we obtain the same solution up to a constant.
\vskip0.1true in
\noindent\textbf{Example 3:}
Consider the following system
\begin{align}\label{39}
\phi_{1}' =& 6\phi_{1} - \phi_{2} \notag \\
\phi_{2}' =& 5\phi_{1} + 4\phi_{2}.
\end{align}
we have, using (\ref{34}) that $\alpha = \frac{-1 +2\imath}{5}$ where $\imath= \sqrt{-1}$.  The general solution, using (\ref{33}) is given by
\begin{align}\label{40}
\phi_{1} =& (\frac{1-2\imath}{4})C_{1}e^{(5+2\imath)x}  + C_{2}(\frac{1-2\imath}{5})e^{(5-2\imath)x} \notag \\
\phi_{2} =& -\frac{5\imath}{4}C_{1}e^{(5+2\imath)x} + C_{2}e^{(5-2\imath)x}.
\end{align}
which is again agree (up to a constant) with the exact solution obtain by the standard method.

\section{A criterion for polynomial solutions}
In Ref. \cite{hakan2}, Saad et al give a sufficient and necessary condition for a second-order linear homogeneous differential equation to have a polynomial solution. Although a similar criterion for the existence of a polynomial solution of a linear system such as (\ref{3}) is not possible in general, the following theorem gives a necessary condition.

\begin{theorem}
If $\phi_1(x)$ and $\phi_2(x)$ are $n-$degree polynomial solutions of a linear system  with variable coefficients (\ref{3}), then
\begin{equation}\label{41}
\eta_n(x)\equiv s_n(x)\omega_n(x)-\lambda_n(x) \rho_n(x)=0, \quad\quad n=1,2,\dots
\end{equation}
where $\lambda_n,s_n,\omega_n$ and $\rho_n$ are given recursively by (\ref{25}).
\end{theorem}
\vskip 0.1 true in
\noindent{\bf Proof:} The existence of nontrivial solutions $\phi_1$ and $\phi_2$ of (\ref{6}) required the vanishing of the determinant 
\begin{equation}\label{42}
\left| \begin{array}{ll}
  \lambda_n(x) & s_n(x) \\
  \omega_n(x) & \rho_n(x)  
  \end{array}\right| = 0.
\end{equation}
 \qed

\noindent Note that the converse of this theorem is not true in general, for example, consider the system $\phi_1^\prime=5x(\phi_1+\phi_2), \phi_2^\prime=3x(\phi_1+\phi_2)$, clearly $\eta_1(x)=0$, however the system has a non-polynomial general solution given by $\phi_1={5\over 8}C_1e^{4x^2}-C_2,$ and $\phi_2={3\over 8}C_1e^{4x^2}+C_2$.

\section{Elementary systems of differential equations}
\noindent By means of the iteration sequence (\ref{25}), it is interesting to note that the termination condition (\ref{22}) can be written, equivalently, as Riccati equation,
\begin{equation} \label{43}
\left( \frac{\lambda_{n}}{s_{n}} \right)'+(\lambda_0-\rho_0)\left(\frac{\lambda_{n}}{s_{n}}\right)-s_{0}\left(\frac{\lambda_{n}}{s_{n}}\right)^2=-w_0
\end{equation}
where the solution of this equation yields the exact analytic expression of the ratio $\alpha\equiv {s_n\over \lambda_n}$ by which the exact solution of the corresponding system follows immediately using (\ref{21}). Thus, we have a useful connection between the exact solution of a linear-system of differential equations and the exact solution of a Riccati equation; namely,
\vskip 0.1 true in
\begin{theorem} The general solution of the linear system:
\begin{align} 
\phi_{1}' =& \lambda_{0}(x)\phi_{1} + s_{0}(x)\phi_{2} \notag \\ 
\phi_{2}' =& \omega_{0}(x)\phi_{1} + \rho_{0}(x)\phi_{2} \label{44}
\end{align}
is given by:
\begin{align} 
\phi_{2}(x) =& \exp\left( \int^{x} \left( \rho_{0}(t) - \alpha \omega_{0}(t)\right) dt \right) \left[ C_{2} + C_{1} \int^{x} \omega_{0}(t) \exp\left( \int^{\tau} \left( \lambda_{0}({\tau}) - \rho_{0}({\tau}) + 2\omega_{0}({\tau})\alpha \right) d\tau \right) dt \right] \notag \\
\phi_{1}(x) =& C_{1}\exp \left(\int^{x}\left(\alpha\omega_{0}(t) + \lambda_{0}(t) \right) dt \right) - \alpha \phi_{2}(x) \label{45}
\end{align}
where $\alpha\equiv \alpha(x)$ is the solution of the Riccati equation
\begin{equation} \label{46}
{d\alpha\over dx}=w_0\alpha^2+(\lambda_0-\rho_0)\alpha-s_0.
\end{equation}
\end{theorem}
Clearly, for linear constant-coefficient systems, (\ref{46}) is equivalent to (\ref{34}) because ${d\alpha/ dx}=0$. Recently, Saad et al \cite{hakan4} studied the exact solutions of certain classes of such Riccati equation which allow us to obtain exact solutions to many systems of linear differential equations. We, first, consider the simpler case, namely, the first iteration using Theorem 1, we obtain:

\begin{theorem}
For the following system of differential equations:
\begin{align} 
\phi_{1}' =& \lambda_{0}(x)\phi_{1} + s_{0}(x)\phi_{2}, \notag \\ 
\phi_{2}' =& \omega_{0}(x)\phi_{1}+\rho_{0}(x)\phi_{2}, \label{47}
\end{align}
if the functions $\lambda_{0}$, $s_{0}$, $\omega_{0}$, and $\rho_{0}$ satisfy:
\begin{equation} \label{48}
\left(\frac{\lambda_{0}}{s_{0}}\right)^\prime - \left(\frac{\lambda_{0}}{s_{0}}\right)\rho_{0} = -\omega_{0}
\end{equation}
then the general solution to the system is given by:
\begin{align} 
\phi_{2} =& \frac{\lambda_{0}(x)}{s_{0}(x)}
\left[C_1\int^x 
{w_0(t)s_0(t)\over \lambda_0(t)} \exp\left({\int^t \left(\lambda_0(\tau)+{s_0(\tau)w_0(\tau)\over \lambda_0(\tau)}\right)d\tau}\right)dt +C_2 \right] \notag \\ 
\phi_{1} =& C_1\exp \left( \int^x \left(\lambda_{0}(t) +{s_0(t)w_0(t)\over \lambda_0(t)}\right)dt \right) -{s_0(x)\over \lambda_0(x)}\phi_{2} \label{49}
\end{align}
\end{theorem}
\vskip0.1true in
\noindent{Proof:} Equation (\ref{46}) can be written as
\begin{equation}\label{50}
{d\over dx}\left(-{1\over \alpha}\right)=w_0+{(\lambda_0-\rho_0)\over \alpha}-{s_0\over \alpha^2}
\end{equation}
For $n=0$, $\alpha\equiv {s_0\over \lambda_0}$, and (\ref{50}) reduces to (\ref{48}).\qed
\vskip0.1 true in
\noindent As direct examples of Theorem 5, we have the following two corollaries.
\begin{corollary}
The following system of differential equations:
\begin{align}
\phi_{1}' &= f(x)\left(\phi_{1} + \phi_{2} \right) \notag \\ 
\phi_{2}' &= g(x)\left(\phi_{1}+\phi_{2} \right)  \label{51}
\end{align}
where $f(x)$ and $g(x)$ are arbitrary functions, has the general solution given by:
\begin{align}
\phi_{1} &= C_{1}\left[\exp \left( \int^x \left(f(t) +g(t) \right)dt \right) -\int^x \left( g(t)\exp \left( \int^t \left(f(\tau) +g(\tau) \right)d\tau \right) \right)dt \right] - C_{2}, \notag \\ 
\phi_{2} &= C_{1}\int^x \left( g(t)\exp \left( \int^t \left(f(\tau) +g(\tau) \right)d\tau \right) \right)dt + C_{2}. \label{52}
\end{align}
\end{corollary}

\begin{corollary}
The following system of differential equations:
\begin{align} 
\phi_{1}' &= f(x)\left(\phi_{1} - \phi_{2} \right) \notag \\ \label{53}
\phi_{2}' &= g(x)\left(\phi_{1} - \phi_{2} \right)
\end{align}
where $f(x)$ and $g(x)$ are arbitrary functions, has the general solution:
\begin{align} 
\phi_{1} &= C_{1}\left[\exp \left( \int^{x} \left(f(t)-g(t)\right)dt \right)+\int^{x}\left( g(x)\exp \left( \int^{t}\left(f(\tau)-g(\tau) \right) d\tau \right) \right)dt\right] - C_{2} \notag \\ 
\phi_{2} &= C_{1}\int^{x}\left( g(t)\exp \left( \int^{t}\left(f(\tau)-g(\tau) \right) d\tau \right) \right)dt - C_{2} \label{54}
\end{align}
\end{corollary}
\vskip0.1true in
\noindent Using theorem (4), we can now prove the following.
\begin{theorem}
The system of differential equations:
\begin{align} 
\phi_{1}' =& \lambda_{0}(x)\phi_{1} + s_{0}(x)\phi_{2}, \notag \\ 
\phi_{2}' =& s_{0}(x)\phi_{1}+\rho_{0}(x)\phi_{2}, \label{54a}
\end{align}
where the functions $\lambda_{0}$, $s_{0}$,  and $\rho_{0}$ are assumed to be continuous and differentiable  functions, is analytically solvable if the quantity $(\rho(x)-\lambda_0(x))/s_0(x)$ is independent of $x$.
\end{theorem}

\noindent{Proof:} If the quantity $(\rho(x)-\lambda_0(x))/s_0(x)$ is independent of $x$, then the Riccati equation (\ref{46}) becomes a separable equation.\qed

\begin{table}
\begin{center} 
\caption{Exact solutions of the system $\phi'_1=\lambda_0(x)\phi_1+s_0(x)\phi_2,~~\phi'_2=w_0(x)\phi_1+\rho_0(x)\phi_2$ for certain functions $\lambda_0(x), s_0(x),w_0(x),$ and $\rho_0(x)$. } 
\begin{tabular}{|l|l|l|l|l|} 
\hline
\hline 
 $\lambda_0(x)$ & $s_0(x)$ & $\omega_0(x)$ & $\rho_0(x)$ & Solution \\ [0.5ex]
\hline\hline 
 $f(x)$ & $g(x)$ & $g(x)$ & $f(x)$ & $\phi_1=C_1e^{\int^x({g(t)\alpha(t)}+f(t))dt}-\alpha(x)\phi_2,$ \\ 
~ &~& ~ &~&$\phi_2=e^{\int(f(t)-g(t)\alpha(t))dt}[C_2+C_1\int^xg(t)e^{\int^t({2g(\tau)\alpha(\tau)})d\tau}dt],$\\[1ex]
~ &~& ~ &~&where $\alpha(x)=\tanh(-\int^x g(t)dt).$\\[1ex]
\hline 
 $f(x)$ & $ag(x)$ & $bg(x)$ & $f(x)$ & $\phi_1=C_1e^{\int^x({bg(t)\alpha(t)}+f(t))dt}-\alpha(x)\phi_2,$ \\ 
~ &~& ~ &~&$\phi_2=e^{\int(f(t)-bg(t)\alpha(t))dt}[C_2+C_1\int^xbg(t)e^{\int^t({2bg(\tau)\alpha(\tau)})d\tau}dt],$\\[1ex]
~ &~& ~ &~&where $\alpha(x)=-\sqrt{a\over b}\tanh(\sqrt{ab}\int^x g(t)dt).$\\[1ex]
\hline 
 $f(x)$ & $-g(x)$ & $g(x)$ & $f(x)$ & $\phi_1=C_1e^{\int^x({g(t)\alpha(t)}+f(t))dt}-\alpha(x)\phi_2,$ \\ 
~ &~& ~ &~&$\phi_2=e^{\int(f(t)-g(t)\alpha(t))dt}[C_2+C_1\int^xg(t)e^{\int^t({2g(\tau)\alpha(\tau)})d\tau}dt],$\\[1ex]
~ &~& ~ &~&where $\alpha(x)=\tan(\int^x g(t)dt).$\\[1ex]
\hline 
 $f(x)$ & $-ag(x)$ & $bg(x)$ & $f(x)$ & $\phi_1=C_1e^{\int^x({bg(t)\alpha(t)}+f(t))dt}-\alpha(x)\phi_2,$ \\ 
~ &~& ~ &~&$\phi_2=e^{\int(f(t)-bg(t)\alpha(t))dt}[C_2+C_1\int^xbg(t)e^{\int^t({2bg(\tau)\alpha(\tau)})d\tau}dt],$\\[1ex]
~ &~& ~ &~&where $\alpha(x)=\sqrt{a\over b}\tan(\sqrt{ab}\int^x g(t)dt).$\\[1ex]
\hline
 $\lambda_0(x)$ & $s_0(x)$ & $\omega_0(x)$ & ${d\over dx}\left({\rho_0-\lambda_0\over s_0}\right)$ & $\phi_1=C_1e^{\int^x({s_0w_0\over \lambda_0-\rho_0}+\lambda_0)dt}-({\lambda_0-\rho_0\over s_0})\phi_2,$ \\ 
~ &~& ~ &~&$\phi_2=e^{\int(\rho_0-{w_0s_0\over \lambda_0-\rho_0})dt}[C_2+C_1\int^xw_0e^{\int^t(\lambda_0-\rho_0+{2w_0s_0\over \lambda_0-\rho_0})d\tau}dt].$\\[1ex]
\hline 
 $\lambda_0(x)$ & ${d\over dx}({\lambda_0-\rho_0\over w_0})$ & $\omega_0(x)$ & $\rho_0(x)$&$\phi_1=C_1e^{\int^x\rho_0(t)dt}+({\lambda_0-\rho_0\over w_0})\phi_2,$ \\  
~ &~& ~ &~&$\phi_2=e^{\int\lambda_0 dt}[C_2+C_1\int^xw_0e^{\int^t(\rho_0-\lambda_0)d\tau}dt].$\\[1ex]
\hline 
\end{tabular}
\end{center}
\end{table}

\vskip0.1true in
\begin{theorem}
The system of differential equations:
\begin{align} 
\phi_{1}' =& \lambda_{0}(x)\phi_{1} + s_{0}(x)\phi_{2}, \notag \\ 
\phi_{2}' =& w_{0}(x)\phi_{1}+\rho_{0}(x)\phi_{2}, \label{54b}
\end{align}
where the functions $\lambda_{0}$, $s_{0}$, $w_0$  and $\rho_{0}$ are assumed to be continuous and differentiable  functions, is analytically solvable if  the quantities $w_0(x)/(\rho(x)-\lambda_0(x))$ and $s_0(x)/(\rho(x)-\lambda_0(x))$ are independent of $x$.
\end{theorem}

\noindent{Proof:} If the quantities $w_0(x)/(\rho(x)-\lambda_0(x))$ and $s_0(x)/(\rho(x)-\lambda_0(x))$ are independent of $x$, then the Riccati equation (\ref{46}) becomes a separable equation.\qed

\section{Solvable linear systems of differential equations}
\noindent For higher iteration levels, $n=1,2,\dots$, the results of the earlier work of Saad et all \cite{hakan4} on Riccati equation \cite{Kurilin}-\cite{Rainville} can be used to obtain exact analytical solutions to different classes of the linear system (\ref{3}). Under a certain conditions on the functions $\lambda_0,s_0,w_0$ and $\rho_0$, we may express \cite{slater} some of these exact solutions in terms of generalized hypergeometric functions ${}_qF_p(\alpha_1,\dots,\alpha_p;\beta_1,\dots,\beta_q;x)$ 
\begin{equation}\label{55}
{}_pF_q(\alpha_1,\alpha_2,\dots,\alpha_p;\beta_1,\beta_2,\dots,\beta_q;x)=\sum\limits_{k=0}^\infty {(\alpha_1)_k(\alpha_2)_k \dots(\alpha_p)_k\over (\beta_1)_k(\beta_2)_k\dots(\beta_q)_k}{x^k\over k!}
\end{equation}
where $p$ and $q$ are nonnegative integers and no $\beta_k,k=1,2,\dots,q$ is zero or a negative integer. Clearly, (\ref{55}) includes the special cases of the confluent hypergeometric function ${}_1F_1$ and the classical `Gaussian' hypergeometric function ${}_2F_1$. The Pochhammer symbol $(\alpha)_k$ is defined in terms of Gamma function as
\begin{equation}\label{56}
(\alpha)_k={\Gamma(\alpha+n)\over \Gamma(\alpha)}\quad k=0,1,2,\dots
\end{equation}
If $\alpha$ is a negative integer $-n$, we have
\begin{equation}\label{57}
(-n)_k=\begin{cases} {(-1)^k n!\over (n-k)!}& 0\leq k\leq n\\
0& k>n
\end{cases}
\end{equation}
in which case, the generalized hypergeometric series reduces to a polynomial of degree $n$ in its variable $x$. In Table IV, we report the exact expression of $\alpha$ for a given system, by which we can easily obtain the exact solutions through a direct substitution of $\alpha$ in (\ref{21}). 

\begin{table}
\begin{center} 
\caption{Exact expressions of the ratio $\alpha\equiv {s_n\over \lambda_n}$, $n=1,2,\dots$ for the system $\phi'_1=\lambda_0(x)\phi_1+s_0(x)\phi_2,\phi'_2=w_0(x)\phi_1+\rho_0(x)\phi_2$, for different $\lambda_0(x),s_0(x),\omega_0(x)$ and $\rho_0(x)$. The exact solutions of the system then follows by direct substitution of $\alpha$ in (\ref{45}). Here $R(x)$ is an arbitrary differentiable function.} 
\begin{tabular}{|l|l|l|l|l|} 
\hline\hline 
 $\lambda_0(x)$ & $s_0(x)$ & $\omega_0(x)$ & $\rho_0(x)$ & $\alpha\equiv{s_n\over \lambda_n}$ \\ [0.5ex]
\hline 
${R'(x)\over R(x)}$ & ${4n\over R(x)}-\left({R'(x)\over R^2(x)}-{2x\over R(x)}\right)^\prime$  & $-R(x)$ & $2x$  &  ${1\over R(x)}\left(-4nx~{{}_1F_1(-n+1;{3\over 2};x^2)
\over {}_1F_1(-n;{1\over 2};x^2)}
+{R'(x)\over R(x)}-2x\right)$\\[1ex]
\hline 
${R'(x)\over R(x)}$ & ${2(2n+1)\over R(x)}-\left({R'(x)\over R^2(x)}-{2x\over R(x)}\right)^\prime$    & $-R(x)$ & $2x$ & ${1\over R(x)}\left({1\over x}-{4nx\over 3}~{{}_1F_1(-n+1;{5\over 2};x^2)
\over {}_1F_1(-n;{3\over 2};x^2)}+{R'(x)\over R(x)}-2x\right)$\\[1ex]
\hline
${R'(x)\over R(x)}$ & ${2na\over R(x)}-\left({R'(x)\over R^2(x)}-{(ax+b)\over R(x)}\right)^\prime$    & $-R(x)$ & $ax+b$ & ${1\over R(x)}\left(-{2n(ax+b)}~{{}_1F_1(-n+1;{3\over 2};{(ax+b)^2\over 2a})
\over {}_1F_1(-n;{1\over 2};{(ax+b)^2\over 2a})}+{R'(x)\over R(x)}-ax-b\right)$\\[1ex]
\hline 
${R'(x)\over R(x)}$ & ${(2n+1)a\over R(x)}-\left({R'(x)\over R^2(x)}-{(ax+b)\over R(x)}\right)^\prime$    & $-R(x)$ & $ax+b$ & ${1\over R(x)}\left({a\over ax+b}-{2n(ax+b)\over 3}~{{}_1F_1(-n+1;{5\over 2};{(ax+b)^2\over 2a})
\over {}_1F_1(-n;{3\over 2};{(ax+b)^2\over 2a})}+{R'(x)\over R(x)}-ax-b\right)$\\[1ex]
\hline 
${R'(x)\over R(x)}$ & ${nb\over xR(x)}-\left({R'(x)\over R^2(x)}-{(b-{c\over x})\over R(x)}\right)^\prime$    & $-R(x)$ & $b-{c\over x}$ & ${1\over R(x)}\left(-{nb\over c}~{{}_1F_1(-n+1;{c+1};{bx})
\over {}_1F_1(-n;c;bx)}+{R'(x)\over R(x)}+{c\over x}-b\right)$\\[1ex]
\hline 
${R'(x)\over R(x)}$ & $-{n^2\over x(1-x)R(x)}-\left({R'(x)\over R^2(x)}-{(-2n+1)x-c\over x(1-x)R(x)}\right)^\prime$    & $-R(x)$ & ${(-2n+1)x-c\over x(1-x)}$ & ${1\over R(x)}\left({n^2\over c}~{{}_2F_1(-n+1,-n+1;{c+1};{x})
\over {}_2F_1(-n,-n;c;x)}+{R'(x)\over R(x)}-{(-2n+1)x-c\over x(1-x)}\right)$\\[1ex]
\hline 
${R'(x)\over R(x)}$ & ${nb\over x(1-x)R(x)}-\left({R'(x)\over R^2(x)}-{(-n+b+1)x-c\over x(1-x)R(x)}\right)^\prime$    & $-R(x)$ & ${(-n+b+1)x-c\over x(1-x)}$ & ${1\over R(x)}\left(-{n b\over c}~{{}_2F_1(-n+1,b+1;{c+1};{x})
\over {}_2F_1(-n,b;c;x)}+{R'(x)\over R(x)}-{(-n+b+1)x-c\over x(1-x)}\right)$\\[1ex]
\hline 
${R'(x)\over R(x)}$ & $-{n(n+a-1)\over x^2R(x)}-\left({R'(x)\over R^2(x)}-{(ax+b)\over x^2R(x)}\right)^\prime$    & $-R(x)$ & $-{ax+b\over x^2}$ & ${1\over R(x)}\left({n(n+a-1)\over b}~{{}_2F_0(-n+1,n+a;-;-{x\over b})
\over {}_2F_0(-n,n+a-1;-;-{x\over b})}+{R'(x)\over R(x)}+{(ax+b)\over x^2}\right)$\\[1ex]
\hline 
${R'(x)\over R(x)}$ & ${n(n+2k+1)\over (1-x^2)R(x)}-\left({R'(x)\over R^2(x)}-{2(k+1)x\over (1-x^2)R(x)}\right)^\prime$    & $-R(x)$ & ${2(k+1)x\over (1-x^2)}$ & ${1\over R(x)}\left({n(n+2k+1)\over 2(k+1)}~{{}_2F_1(-n+1,n+2k+2;k+2;{(1-x)\over 2})
\over {}_2F_1(-n,n+2k+1;k+1;{(1-x)\over 2})}+{R'(x)\over R(x)}-{2(k+1)x\over (1-x^2)}\right)$\\[1ex]
\hline 
${R'(x)\over R(x)}$ & ${n(n+2k)\over (1-x^2)R(x)}-\left({R'(x)\over R^2(x)}-{(2k+1)x\over (1-x^2)R(x)}\right)^\prime$    & $-R(x)$ & ${(2k+1)x\over (1-x^2)}$ & ${1\over R(x)}\left({n(n+2k)\over (2k+1)}~{{}_2F_1(-n+1,n+2k+1;k+{3\over 2};{(1-x)\over 2})
\over {}_2F_1(-n,n+2k;k+{1\over 2};{(1-x)\over 2})}+{R'(x)\over R(x)}-{(2k+1)x\over (1-x^2)}\right)$\\[1ex]
\hline 
\end{tabular}
\end{center}
\end{table}

\begin{theorem}\label{th6}
If functions $F_0\equiv F_{0}(x)$ and $G_0\equiv G_{0}(x)$ satisfy the recursive relation:
\begin{equation}\label{57}
\delta_n\equiv F_{n}G_{n-1} - F_{n-1}G_{n} = 0,\quad\mbox{for some}\quad \quad n=1,2,\dots
\end{equation}
where 
\begin{equation}\label{58}
\begin{cases} {F_{n} = F_{n-1}'+G_{n-1}+F_{0}F_{n-1}},\\
G_{n} = G_{n-1}' + G_{0}F_{n-1},
\end{cases}
\end{equation}
then the linear system of differential equations:
\begin{align}  \label{59} 
\phi_{1}' =&  \lambda_{0}\phi_{1} + G_{0}\exp\left(-\int^x (F_{0}+ \rho_{0}-\lambda_{0}) dt \right)\phi_{2} \notag \\
\phi_{2}' =& \exp\left(\int^x( F_{0} + \rho_{0}-\lambda_{0})dt\right)\phi_{1} + \rho_{0}\phi_{2}
\end{align}
has a general solution given by (\ref{45}) with $s_0= G_{0}\exp\left(-\int^x (F_{0}+ \rho_{0}-\lambda_{0}) dt \right)$, $w_0=\exp\left(\int^x( F_{0} + \rho_{0}-\lambda_{0})dt\right)$, and
\begin{equation}\label{60}
\alpha ={G_{n-1}\over F_{n-1}} \exp\left(-\int( F_{0} + \rho_{0}-\lambda_{0})dx\right),
\end{equation}
where $\lambda_0\equiv \lambda_0(x)$ and $s_0\equiv s_0(x)$ are arbitrary functions.
\end{theorem}
\vskip0.1true in
\noindent{Proof:} The substitution $\alpha=-{u'\over w_0u}$ transfer the Riccati equation (\ref{46}) into the second-order differential equation
\begin{equation}\label{61}
u''=\left[{w_0^\prime\over w_0}+\lambda_0-\rho_0\right]u'+s_0w_0u
\end{equation}
Using the standard theorem of the asymptotic iteration method \cite{hakan1}, the solution of (\ref{61}) satisfies \cite{hakan3}
\begin{equation}\label{62}
{u'\over u}=-{G_{n-1}\over F_{n-1}}
\end{equation}
if, for some $n>0$,
\begin{equation*}
\delta_n\equiv F_{n}G_{n-1} - F_{n-1}G_{n} = 0,
\end{equation*}
where $F_n$ and $G_n$ are computed using the recursive relations
\begin{equation*}
\begin{cases} {F_{n} = F_{n-1}'+G_{n-1}+F_{0}F_{n-1}},\\
G_{n} = G_{n-1}' + G_{0}F_{n-1}.
\end{cases}
\end{equation*}
Thus, the general solution of (\ref{59}), using Theorem 4, is given by (\ref{45}) with
\begin{equation}\label{63}
\alpha ={G_{n-1}\over w_0F_{n-1}},
\end{equation}
which complete the prove of the theorem.\qed
\vskip0.1true in
\noindent{\bf Example 4:} Let $F_0=2x$ and $G_0=-2n,\quad n=1,2\dots$. Using (\ref{58}), see also \cite{hakan3}, we know that $\delta_n=F_{n}G_{n-1} - F_{n-1}G_{n}=0$ for $n=1,2,\dots$ with
\begin{equation}\label{64}
{G_{n-1}\over F_{n-1}}=-{1\over H_{n}(x)}{dH_n(x)\over dx}
\end{equation}
where $H_n(x)$ is the well-known Hermite polynomials. Thus, for $\rho_0\equiv \rho_0(x)$ and $\lambda_0\equiv 
\lambda_0(x)$ arbitrary functions, the system 
\begin{align}
\phi_{1}' =&  \lambda_{0}\phi_{1} -2n\exp\left(-\int^x (2x+ \rho_{0}-\lambda_{0}) dt \right)\phi_{2} \notag \\
\phi_{2}' =& \exp\left(\int^x(2x+ \rho_{0}-\lambda_{0})dt\right)\phi_{1} + \rho_{0}\phi_{2}  \label{59} 
\end{align}
has a general solution
\begin{align} 
\phi_{2}(x) =& H_n(x)e^{\int^{x} \rho_{0}(t)dt} 
\left[ C_{2} + C_{1} \int^{x}{e^{t^2}\over [H_n(t)]^2}dt \right], \notag \\
\phi_{1}(x) =& {C_{1}\over H_n(x)}e^{\int^{x}\lambda_{0}(t)dt}- \alpha \phi_{2}(x). \label{65}
\end{align}
In Table V, we gave the exact expressions of $\alpha\equiv {G_{n-1}\over w_0(x)\lambda_{n-1}}$, $n=1,2,\dots$ for different systems. The general solutions can be found by direct substitution of $\alpha$ in (\ref{45}). 

\begin{table}
\begin{center} 
\caption{Exact expressions of the ratio $\alpha\equiv {G_{n-1}\over w_0(x)F_{n-1}}$, $n=1,2,\dots$ for the system $\phi'_1=\lambda_0(x)\phi_1+s_0(x)\phi_2,~~\phi'_2=w_0(x)\phi_1+\rho_0(x)\phi_2$, for arbitrary functions $\lambda_0(x)$ and $\rho_0(x)$. The exact solutions of the system then follows by direct substitution of $\alpha$ in (\ref{45}). } 
\begin{tabular}{|l|l|l|} 
\hline\hline 
 $s_0(x)$ & $\omega_0(x)$ &  $\alpha\equiv{1\over \omega_0(x)}{G_{n-1}\over F_{n-1}}$ \\ [1ex]
\hline 
 $-2ne^{-\int^x\left(2x+\rho_0-\lambda_0\right)dx}$  & $e^{\int^x\left(2x+\rho_0-\lambda_0\right)dx}$ &   $-{{d\over dx}(H_{n}(x))\over H_{n}(x)}e^{-\int^x\left(2x+\rho_0-\lambda_0\right)dx} $\\[1ex]
\hline 
$-ane^{-\int^x\left(ax+b+\rho_0-\lambda_0\right)dx},~n=2,4,6,\dots$  & $e^{\int^x\left(ax+b+\rho_0-\lambda_0\right)dx}$ &  $-{{d\over dx}\left({}_1F_1(-{n\over 2};{1\over 2};{(ax+b)^2\over 2a})\right)\over {}_1F_1(-{n\over 2};{1\over 2};{(ax+b)^2\over 2a})}e^{-\int^x\left(ax+b+\rho_0-\lambda_0\right)dx},$\\[1ex]
\hline
 $-ane^{-\int^x\left(ax+b+\rho_0-\lambda_0\right)dx},~n=1,3,5,\dots$  & $e^{\int^x\left(ax+b+\rho_0-\lambda_0\right)dx}$ &  $-{{d\over dx}\left((ax+b){}_1F_1(-{n-1\over 2};{3\over 2};{(ax+b)^2\over 2a})\right)\over (ax+b){}_1F_1(-{n-1\over 2};{3\over 2};{(ax+b)^2\over 2a})}e^{-\int^x\left(ax+b+\rho_0-\lambda_0\right)dx},$\\[1ex]
\hline 
 $-{bn\over x}e^{-\int^x\left(b-{c\over x}+\rho_0-\lambda_0\right)dx}$  & $e^{\int^x\left(b-{c\over x}+\rho_0-\lambda_0\right)dx}$ &  $-{{d\over dx}\left[{}_1F_1(-n;c;bx)\right]\over {}_1F_1(-n;c;bx)}e^{-\int^x\left(b-{c\over x}+\rho_0-\lambda_0\right)dx}$\\[1ex]
\hline 
${n^2\over x(1-x)}e^{-\int^x\left({(-2n+1)x-c\over x(1-x)}+\rho_0-\lambda_0\right)dx}$  & $e^{\int^x\left({(-2n+1)x-c\over x(1-x)}+\rho_0-\lambda_0\right)dx}$ &  $-{{d\over dx}\left[{}_2F_1(-n,-n;c;x)\right]\over {}_1F_1(-n,-n;c;x)}e^{-\int^x\left({(-2n+1)x-c\over x(1-x)}+\rho_0-\lambda_0\right)dx}$\\[1ex]
\hline 
 $-{n(n+1)\over (1-x^2)}e^{-\int^x\left({2x\over (1-x^2)}+\rho_0-\lambda_0\right)dx}$  & $e^{\int^x\left({2x\over (1-x^2)}+\rho_0-\lambda_0\right)dx}$  &  $-{{d\over dx}\left[{}_2F_1(-n,n+1;1;{(1-x)\over 2})\right]\over {}_1F_1(-n,n+1;1;{(1-x)\over 2})}e^{-\int^x\left({2x\over (1-x^2)}+\rho_0-\lambda_0\right)dx}$\\[1ex]
\hline 
 $-{n(n+a+b+1)\over (1-x^2)}e^{-\int^x\left({(a+b+2)x+a-b\over (1-x^2)}+\rho_0-\lambda_0\right)dx}$  & $e^{\int^x\left({(a+b+2)x+a-b\over (1-x^2)}+\rho_0-\lambda_0\right)dx}$  &  $-{{d\over dx}\left[{}_2F_1(-n,n+a+b+1;a+1;{(1-x)\over 2})\right]\over {}_2F_1(-n,n+a+b+1;a+1;{(1-x)\over 2})}e^{-\int^x\left({(a+b+2)x+a-b\over (1-x^2)}+\rho_0-\lambda_0\right)dx}$\\[1ex]
\hline 
 $-{n(n+2)\over (1-x^2)}e^{-\int^x\left({3x\over 1-x^2}+\rho_0-\lambda_0\right)dx}$  & $e^{\int^x\left({3x\over 1-x^2}+\rho_0-\lambda_0\right)dx}$  &  $-{{d\over dx}\left[{}_2F_1(-n,n+2;{3\over 2};{(1-x)\over 2})\right]\over {}_2F_1(-n,n+2;{3\over 2};{(1-x)\over 2})}e^{-\int^x\left({3x\over 1-x^2}+\rho_0-\lambda_0\right)dx}$\\[1ex]
\hline 
 $-{n(n+2k)\over (1-x^2)}e^{-\int^x\left({(1+2k)x\over 1-x^2}+\rho_0-\lambda_0\right)dx}$  & $e^{\int^x\left({(1+2k)x\over 1-x^2}+\rho_0-\lambda_0\right)dx}$  &  $-{{d\over dx}\left[{}_2F_1(-n,n+2k;k+{1\over 2};{(1-x)\over 2})\right]\over {}_2F_1(-n,n+2k;k+{1\over 2};{(1-x)\over 2})}e^{-\int^x\left({(1+2k)x\over 1-x^2}+\rho_0-\lambda_0\right)dx}$\\[1ex]
\hline 
 $-{n(n+2k+1)\over (1-x^2)}e^{-\int^x\left({2(1+k)x\over 1-x^2}+\rho_0-\lambda_0\right)dx}$  & $e^{\int^x\left({2(1+k)x\over 1-x^2}+\rho_0-\lambda_0\right)dx}$  &  $-{{d\over dx}\left[{}_2F_1(-n,n+2k+1;k+{1};{(1-x)\over 2})\right]\over {}_2F_1(-n,n+2k+1;k+{1};{(1-x)\over 2})}e^{-\int^x\left({2(1+k)x\over 1-x^2}+\rho_0-\lambda_0\right)dx}$\\[1ex]
\hline 
 ${n(n+1)\over x^2}e^{-\int^x\left({-2(1+x)\over x^2}+\rho_0-\lambda_0\right)dx}$  & $e^{\int^x\left(-{2(1+x)\over x^2}+\rho_0-\lambda_0\right)dx}$  &  $-{{d\over dx}\left[{}_2F_0(-n,n+1;-;-{x\over 2})\right]\over {}_2F_0(-n,n+1;-;-{x\over 2})}e^{-\int^x\left(-{2(1+x)\over x^2}+\rho_0-\lambda_0\right)dx}$\\[1ex]
\hline 
${n(n+a-1)\over x^2}e^{-\int^x\left({-(ax+b)\over x^2}+\rho_0-\lambda_0\right)dx}$  & $e^{\int^x\left(-{(ax+b)\over x^2}+\rho_0-\lambda_0\right)dx}$  &  $-{{d\over dx}\left[{}_2F_0(-n,n+a-1;-;-{x\over b})\right]\over {}_2F_0(-n,n+a-1;-;-{x\over b})}e^{-\int^x\left(-{(ax+b)\over x^2}+\rho_0-\lambda_0\right)dx}$\\[1ex]
\hline 
\end{tabular}
\end{center}
\end{table}

\begin{theorem}
If functions $F_0\equiv F_{0}(x)$ and $G_0\equiv G_{0}(x)$ satisfy the recursive relation:
\begin{equation}\label{67}
\delta_n\equiv F_{n}G_{n-1} - F_{n-1}G_{n} = 0,\quad\mbox{for some}\quad \quad n=1,2,\dots
\end{equation}
where 
\begin{equation}\label{68}
\begin{cases} {F_{n} = F_{n-1}'+G_{n-1}+F_{0}F_{n-1}},\\
G_{n} = G_{n-1}' + G_{0}F_{n-1},
\end{cases}
\end{equation}
then the linear system of differential equations:
\begin{align}
\phi_{1}' =&  \lambda_{0}\phi_{1} + \exp\left(\int^x (F_{0}- \rho_{0}+\lambda_{0}) dx \right)\phi_{2} \notag \\
\phi_{2}' =& G_0\exp\left(\int^x-( F_{0} - \rho_{0}+\lambda_{0})dx\right)\phi_{1} + \rho_{0}\phi_{2}  \label{69} 
\end{align}
has a general solution given by (\ref{45}) with $s_0= \exp\left(\int^x (F_{0}- \rho_{0}+\lambda_{0}) dx \right)$, $w_0=\exp\left(-\int^x( F_{0} - \rho_{0}+\lambda_{0})dx\right)$, and
\begin{equation}\label{70}
\alpha \equiv{F_{n-1}\over G_{n-1}} \exp\left(\int(F_{0} + \rho_{0}-\lambda_{0})dx\right),
\end{equation}
where $\lambda_0\equiv \lambda_0(x)$ and $s_0\equiv s_0(x)$ are arbitrary functions.
\end{theorem}

\noindent{Proof:} The proof of this theorem follows similarly to the prove of Theorem 6 using the substitution $\alpha=-{s_0u\over u'}$ .\qed
\vskip0.1 true in
\noindent In Table VI, we gave the exact expressions of $\alpha\equiv {s_0F_{n-1}\over G_{n-1}}$, $n=1,2,\dots$ for different systems. The general solutions can be found by direct substitution of $\alpha$ in (\ref{45}). 

\begin{table}
\begin{center} 
\caption{Exact expressions of the ratio $\alpha\equiv {s_0F_{n-1}\over G_{n-1}}$, $n=1,2,\dots$ for the system $\phi'_1=\lambda_0(x)\phi_1+s_0(x)\phi_2,~~\phi'_2=w_0(x)\phi_1+\rho_0(x)\phi_2$, for arbitrary functions $\lambda_0(x)$ and $\rho_0(x)$. The exact solutions of the system then follows by direct substitution of $\alpha$ in (\ref{45}). } 
\begin{tabular}{|l|l|l|} 
\hline\hline 
 $s_0(x)$ & $\omega_0(x)$ &  $\alpha\equiv{s_0F_{n-1}\over G_{n-1}}$ \\ [1ex]
\hline 
 $e^{\int^x\left(2x-\rho_0+\lambda_0\right)dx}$  & $-2ne^{-\int^x\left(2x-\rho_0+\lambda_0\right)dx}$ &   $-{H_{n}(x)\over H^\prime_{n}(x)}e^{\int^x\left(2x+\rho_0-\lambda_0\right)dx} $\\[1ex]
\hline 
$e^{\int^x\left(ax+b-\rho_0+\lambda_0\right)dx}$  & $-ane^{-\int^x\left(ax+b-\rho_0+\lambda_0\right)dx},~n=2,4,6,\dots$ &  $-{{}_1F_1(-{n\over 2};{1\over 2};{(ax+b)^2\over 2a})\over {d\over dx}\left({}_1F_1(-{n\over 2};{1\over 2};{(ax+b)^2\over 2a})\right)}e^{\int^x\left(ax+b-\rho_0+\lambda_0\right)dx}$\\[1ex]
\hline
 $e^{\int^x\left(ax+b-\rho_0+\lambda_0\right)dx},$  & $-ane^{-\int^x\left(ax+b-\rho_0+\lambda_0\right)dx}~n=1,3,5,\dots$ &  $-{(ax+b){}_1F_1(-{n-1\over 2};{3\over 2};{(ax+b)^2\over 2a})\over {{d\over dx}\left((ax+b){}_1F_1(-{n-1\over 2};{3\over 2};{(ax+b)^2\over 2a})\right)}}e^{\int^x\left(ax+b-\rho_0+\lambda_0\right)dx}$\\[1ex]
\hline 
 $e^{\int^x\left(b-{c\over x}-\rho_0+\lambda_0\right)dx}$  & $-{bn\over x}e^{-\int^x\left(b-{c\over x}-\rho_0+\lambda_0\right)dx}$ &  $-{{}_1F_1(-n;c;bx)\over{{d\over dx}\left[{}_1F_1(-n;c;bx)\right] }}e^{\int^x\left(b-{c\over x}-\rho_0+\lambda_0\right)dx}$\\[1ex]
\hline 
$e^{\int^x\left({(-2n+1)x-c\over x(1-x)}-\rho_0+\lambda_0\right)dx}$  & ${n^2\over x(1-x)}e^{-\int^x\left({(-2n+1)x-c\over x(1-x)}-\rho_0+\lambda_0\right)dx}$ &  $-{{}_1F_1(-n,-n;c;x)\over {d\over dx}\left[{}_2F_1(-n,-n;c;x)\right]}e^{\int^x\left({(-2n+1)x-c\over x(1-x)}-\rho_0+\lambda_0\right)dx}$\\[1ex]
\hline 
 $e^{\int^x\left({2x\over (1-x^2)}-\rho_0+\lambda_0\right)dx}$  & $-{n(n+1)\over (1-x^2)}e^{-\int^x\left({2x\over (1-x^2)}-\rho_0+\lambda_0\right)dx}$  &  $-{{}_1F_1(-n,n+1;1;{(1-x)\over 2})\over {d\over dx}\left[{}_2F_1(-n,n+1;1;{(1-x)\over 2})\right]}e^{\int^x\left({2x\over (1-x^2)}-\rho_0+\lambda_0\right)dx}$\\[1ex]
\hline 
 $e^{\int^x\left({(a+b+2)x+a-b\over (1-x^2)}-\rho_0+\lambda_0\right)dx}$  & $-{n(n+a+b+1)\over (1-x^2)}e^{-\int^x\left({(a+b+2)x+a-b\over (1-x^2)}-\rho_0+\lambda_0\right)dx}$  &  $-{{}_2F_1(-n,n+a+b+1;a+1;{(1-x)\over 2})\over {d\over dx}\left[{}_2F_1(-n,n+a+b+1;a+1;{(1-x)\over 2})\right]}e^{\int^x\left({(a+b+2)x+a-b\over (1-x^2)}-\rho_0+\lambda_0\right)dx}$\\[1ex]
\hline 
 $e^{\int^x\left({3x\over 1-x^2}-\rho_0+\lambda_0\right)dx}$  & $-{n(n+2)\over (1-x^2)}e^{-\int^x\left({3x\over 1-x^2}-\rho_0+\lambda_0\right)dx}$  &  $-{{}_2F_1(-n,n+2;{3\over 2};{(1-x)\over 2})\over {d\over dx}\left[{}_2F_1(-n,n+2;{3\over 2};{(1-x)\over 2})\right]}e^{\int^x\left({3x\over 1-x^2}-\rho_0+\lambda_0\right)dx}$\\[1ex]
\hline 
 $e^{\int^x\left({(1+2k)x\over 1-x^2}-\rho_0+\lambda_0\right)dx}$  & $-{n(n+2k)\over (1-x^2)}e^{-\int^x\left({(1+2k)x\over 1-x^2}-\rho_0+\lambda_0\right)dx}$  &  $-{{}_2F_1(-n,n+2k;k+{1\over 2};{(1-x)\over 2})\over {d\over dx}\left[{}_2F_1(-n,n+2k;k+{1\over 2};{(1-x)\over 2})\right]}e^{\int^x\left({(1+2k)x\over 1-x^2}-\rho_0+\lambda_0\right)dx}$\\[1ex]
\hline 
 $e^{\int^x\left({2(1+k)x\over 1-x^2}-\rho_0+\lambda_0\right)dx}$  & $-{n(n+2k+1)\over (1-x^2)}e^{-\int^x\left({2(1+k)x\over 1-x^2}-\rho_0+\lambda_0\right)dx}$  &  $-{{}_2F_1(-n,n+2k+1;k+{1};{(1-x)\over 2})\over{d\over dx}\left[{}_2F_1(-n,n+2k+1;k+{1};{(1-x)\over 2})\right] }e^{\int^x\left({2(1+k)x\over 1-x^2}-\rho_0+\lambda_0\right)dx}$\\[1ex]
\hline 
 $e^{\int^x\left({-2(1+x)\over x^2}-\rho_0+\lambda_0\right)dx}$  & ${n(n+1)\over x^2}e^{-\int^x\left(-{2(1+x)\over x^2}-\rho_0+\lambda_0\right)dx}$  &  $-{{}_2F_0(-n,n+1;-;-{x\over 2})\over {d\over dx}\left[{}_2F_0(-n,n+1;-;-{x\over 2})\right]}e^{\int^x\left(-{2(1+x)\over x^2}-\rho_0+\lambda_0\right)dx}$\\[1ex]
\hline 
$e^{\int^x\left({-(ax+b)\over x^2}-\rho_0+\lambda_0\right)dx}$  & ${n(n+a-1)\over x^2}e^{-\int^x\left(-{(ax+b)\over x^2}-\rho_0+\lambda_0\right)dx}$  &  $-{ {}_2F_0(-n,n+a-1;-;-{x\over b})\over{d\over dx}\left[{}_2F_0(-n,n+a-1;-;-{x\over b})\right]}e^{\int^x\left(-{(ax+b)\over x^2}-\rho_0+\lambda_0\right)dx}$\\[1ex]
\hline 
\end{tabular}
\end{center}
\end{table}
\section{Conclusion}
\noindent Guided by the recent applications  of the asymptotic iteration method on Dirac equation and Riccati equation \cite{hakan1}-\cite{hakan4}, we extend AIM to solve analytically different systems of linear differential equations with variable coefficients. This allow us to provide a simple recipe for solving any linear constant-coefficient systems in few steps. Further, we have extended the standard list of exactly solvable system known in the literature (see \cite{polyanin}, Chapter T6, Section T6.1.1, Page 1229).  Furthermore, the work presented here established a connection between the solution of arbitrary linear system of differential equation and the solution of Riccati equation. Indeed, one of the main results in the present work, is to have this connection written explicitly in terms of a theorem; namely Theorem 4. Another interesting conclusion of the present work is that it provides a set of conditions on $\lambda_0, s_0, w_0,\rho_0$ which govern the complete solvability of the system (\ref{3}). It should be clear the results of our investigation are not limited to the exact solutions reported in tables I-VI but it can easily extended to many other solvable systems.  General speaking, the results obtained here are useful in the sense that they can set up a more rational base to approach to solve coupled differential equations \cite{cao}-\cite{xuan}, provide a means to examine this problem from a qualitative point of view, alleviating then the burden of the computational work.

\bigskip
\section*{Acknowledgments}
\medskip
\noindent The present study was supported, in part, by the {\it Natural Sciences and Engineering Research Council of Canada} under Grant GP600515 (NS).  
\medskip


\end{document}